\documentclass{article}
\usepackage{graphicx} % Required for inserting images
\usepackage{amsmath}
\usepackage{tcolorbox}
\usepackage{colortbl}
\usepackage{algpseudocode}
\usepackage{algorithm}
\usepackage{multirow}
\usepackage{authblk}

\newcommand{\method}{\texttt{IterQR}}

\title{\method{}: An Iterative Framework for LLM-based Query Rewrite in e-Commercial Search System}
\date{}

\author[1]{Shangyu Chen*}
\author[1]{Xinyu Jia\footnote{These authors contributed equally to this work}}
\author[1]{Yingfei Zhang}
\author[1]{Shuai Zhang}
\author[1]{Xiang Li}
\author[1]{Wei Lin}
\affil[1]{Meituan, Beijing, China}
\affil[ ]{\{chenshangyu03,jiaxinyu04,zhangyingfei03\\zhangshuai51,lixiang245,linwei31\}@meituan.com}

\begin{document}

\maketitle

\section{Abstract}
The essence of modern e-Commercial search system lies in matching user's intent and available candidates depending on user's query, providing personalized and precise service.
However, user's query may be incorrect due to ambiguous input and typo, leading to inaccurate search. These cases may be released by query rewrite: modify query to other representation or expansion. 
However, traditional query rewrite replies on static rewrite vocabulary, which is manually established meanwhile lacks interaction with both domain knowledge in e-Commercial system and common knowledge in the real world.
% The emergence of Large Language Models (LLMs) empower opportunity to rephrase query rewrite
In this paper, with the ability to generate text content of Large Language Models (LLMs), we provide an iterative framework to generate query rewrite. 
The framework incorporates a 3-stage procedure in each iteration: 
Rewrite Generation with domain knowledge by Retrieval-Augmented Generation (RAG) and query understanding by Chain-of-Thoughts (CoT); 
Online Signal Collection with automatic positive rewrite update;
Post-training of LLM with multi task objective to generate new rewrites.
Our work (named as \method{}) provides a comprehensive framework to generate \textbf{Q}uery \textbf{R}ewrite with both domain / real-world knowledge. 
It automatically update and self-correct the rewrites during \textbf{iter}ations.
\method{} has been deployed in Meituan Delivery's search system 
 (China's leading food delivery platform), providing service for users with significant improvement.

\section{Introduction}
% Search has become a main entry for user to find interested content, especially in e-Commercial system, such as Taobao, Amazon and Meituan Delivery. These systems contain billions of supply for each search request. Though Meituan focus more on nearby supply based on user's location (Location-based Service, LBS), the huge volume of candidates demand service more than merely returning satisfied content depending on user query, but personalized and most interested candidates by different users, leading to click, transaction with high possibility. 

Search functions as the primary means for users to find desired content, especially in e-Commercial systems like Taobao, Amazon, and Meituan Delivery. These systems host billions of suppliers for each search request. While Meituan focuses more on nearby suppliers based on users' locations (Location-based Services, LBS), the vast number of candidates necessitates a service model that delivers not only satisfactory content but also personalized and relevant suggestions, ultimately enhancing click-through and purchase likelihood.

% However, it is common that user query is redundant, ambiguous and incorrect.  Specially in Meituan Delivery, A query of ``Animal cream cake'' requires both animal cream and cake, which may leads to insufficient retrieval candidates.  Typo commonly appears in query, such as ``wontom'' is a typo of ``wonton''.
% Besides,  query of ``spicy'' is too general to provide precise matching: the exposed content is satisfied for the matching of restaurant's title, missing potential candidates such as Jiangxi cuisine\footnote{Jiangxi cuisine is a type of Chinese cuisine known for its spiciness.}. 
% Abbreviation is commonly used in query, such as ``KFC'' for Kentucky Fried Chicken, ``lsf'' for ``Luosifen'' \footnote{Luosifen, or river snail rice noodle, is a popular Chinese dish originating from Liuzhou in the Guangxi Zhuang Autonomous Region. This dish is known for its unique and strong aroma.}.
% User's query may contain a part of the restaurant's title, such as ``Niujie'' for a barbecue restaurant with title of ``Niujie BBQ''. This query may lead to empty results if the restaurant is not available during the search period.

Nonetheless, it is common for user queries to be redundant, ambiguous, or incorrect. For instance, a query for “Animal cream cake” necessitates identifying both "animal cream" and "cake," which may result in insufficient retrieval candidates. Typos are prevalent in queries, as in ``wontom'', which is a misspelling of ``wonton''. Such errors can severely impact search effectiveness, as the system may not recognize the intended dish.
Moreover, queries that are overly general: such as simply stating ``spicy'', fail to yield precise matches, resulting in exposure primarily associated with generic restaurant titles and missing opportunities for niche cuisines like Jiangxi cuisine\footnote{Jiangxi cuisine is a type of Chinese cuisine known for its spiciness.}. 
Abbreviations are also common in queries, such as ``KFC'' for Kentucky Fried Chicken or ``lsf'' for ``Luosifen''\footnote{Luosifen, or river snail rice noodle, is a popular Chinese dish originating from Liuzhou in the Guangxi Zhuang Autonomous Region. This dish is known for its unique and strong aroma.}. 
Users' queries might include only a part of a restaurant's title, such as ``Niujie'' for a barbecue restaurant named ``Niujie BBQ''. This type of incomplete query can sometimes produce empty search results if the restaurant is unavailable during the search duration, leading to user frustration.

% It is difficult to be addressed comprehensively by traditional retrieval methods such as blurry matching, embedding retrieval and query partition for these cases require both domain knowledge (``Niujie'' is responsible for barbecue), common knowledge (typos of ``wontom'') and query understanding for key word extraction.
Traditional retrieval methods struggle to address these challenges comprehensively. Techniques such as fuzzy matching, embedding retrieval, and query partitioning fail effectively to account for the nuanced dependencies between traditional domain knowledge—like understanding that ``Niujie'' implies barbecue—and common knowledge, such as recognizing typographical errors like ``wontom''. Furthermore, the capability to extract relevant keywords based on query understanding is often lacking in these systems.
% Query rewrite serves as a potential solution. However, prior rewrite methods reply on a static rewrite vocabulary which is established manually: queries with high frequency is prepared with rewrites while the tail queries are seldom processed.
Query rewriting emerges as a viable solution to these problems. It can enhance the clarity and relevance of user queries by transforming them into more precise representations. However, prior methods for query rewriting primarily rely on static rewrite vocabularies that are established manually. High-frequency queries may have pre-prepared rewrites, while low-frequency queries—often consisting of user errors, less common terms, or local specialties—receive little to no attention.
% Though LLMs assistant in rewrite generation automatically \cite{peng2024large},\cite{wang2024one}, these methods fail to answer a key question: how to determine whether a specific rewrite is ``good'' or not. ``Good'' represents the rewrite is able to retrieve both relevance satisfied for query and interested content for user.
Although LLMs facilitate automatic rewrite generation, they do not sufficiently answer a critical question: how can we determine if a specific rewrite is ``good''? A "good" rewrite should effectively retrieve not only relevant content that satisfies user queries but also additional content that interests the user. While offline and online experiments may guide the generation of rewrites, assessing the quality of each rewrite before deployment remains a significant challenge. Moreover, most existing methods generate rewrites just once, lacking mechanisms for updates and improvements.
% Offline and online experiments may provide guidance for a batch of rewrites, it is difficult to distinguish the quality of every rewrite before deployment.
% Besides, most methods generate rewrites only one time, lacking ability for update.

\begin{figure*}
    \centering
    \includegraphics[width=\linewidth]{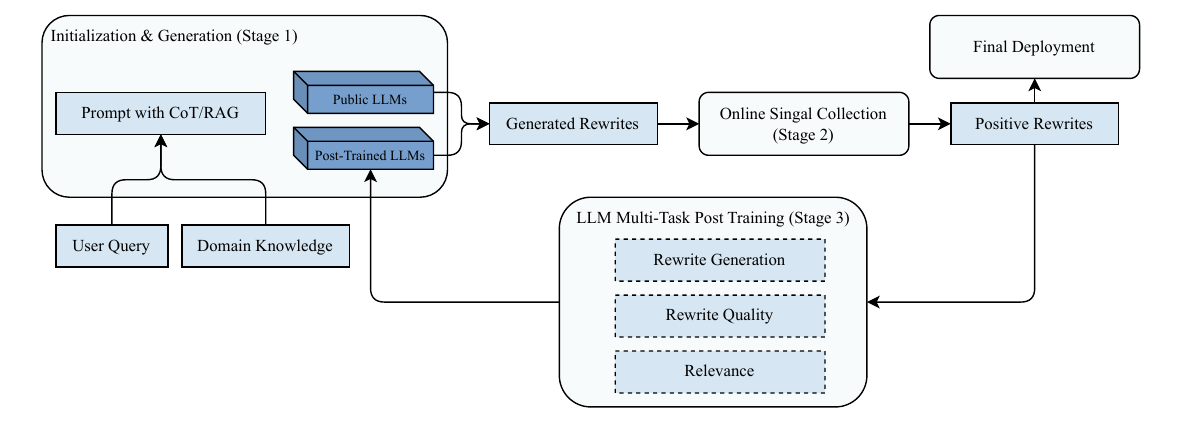}
    \caption{Workflow of \method{}: An iterative framework of 3 stages. Stage 1 initialize and generate query rewrite based on query. Prompt is designed to incorporate domain knowledge (associated interacted restaurant and cuisine. Besides, query rewrite is formulated as query understanding and process by CoT. The generated rewrites are fed to Stage 2 for online feedback collection, where the positive rewrites are utilized in Stage 3. LLM is post trained with multi-task objectives, using the positive rewrites as labels. The trained model serves to motivate new rewrites in Stage 1, prompting the continuous iteration.}
    \label{fig:overall}
\end{figure*}

To address the above mentioned question comprehensively, we proposed \method{}: an iterative framework to generate query rewrites based on LLMs.
As shown in Fig.\ref{fig:overall}: 
% In each iteration, we firstly generate query rewrites using LLMs. Prompt is meticulously designed by rephrasing the task of query rewrite as successive tasks (query understanding, correction, rewrite) using CoT. Then domain knowledge is injected by incorporating interacted title of restaurants and cuisine under the query, instructing LLMs to generate relevant content.
each iteration involves generating query rewrites through LLMs. The prompts for this process are carefully crafted to reformulate the query rewriting task into successive stages (query understanding, correction, rewrite) employing methods like Chain-of-Thoughts (CoT). We inject domain knowledge by incorporating relevant restaurant titles and associated cuisines in the prompts, guiding LLMs to produce pertinent and context-aware content. The generated rewrites feed into our search system for deployment, enhancing user interactions and search effectiveness.
% The generated rewrites are fed into our search system for serving. In the second stage, rewrites leading to interacted items are regarded as positive rewrites and collected. The collection procedure serves as an automatic and precise distinction of the quality of rewrites.
In the subsequent stages, rewrites that lead to user interactions are collected as positive rewrites, serving as indicators of rewrite quality. This automatic and precise distinction of rewrite effectiveness allows us to refine our approach continually. 
% In the last stage, with the collected rewrites, we post-trained a LLM with multi tasks: Rewrite Generation to reproduce the positive rewrites; Rewrite Quality to instruct LLM to distinguish whether the rewrite is ``good'' or not. Relevance to teach LLM with relevance information in the domain of cuisine delivery.
In the final stage, we conduct post-training for the LLM using multi-task objectives to generate new rewrites, including: Rewrite Generation to reproduce the positive rewrites; Rewrite Quality to instruct LLM to distinguish whether the rewrite is ``good'' or not. Relevance to teach LLM with relevance information in the domain of cuisine delivery.

The post-trained LLM is used to generate and motivate new rewrites in first stage, which is then evaluated in second stage. The positive rewrites are filtered and reserved. \method{} provides an iterative framework to continuously generate new rewrites with automatic evaluation. Domain knowledge and new information can be naturally incorporated during generation.

The main contributions of \method{} are listed as follows:
\begin{itemize}
    \item \method{} pioneer to address the automatic evaluation of generated rewrites before final deployment, by feeding rewrites online for signal collection.
    \item An iterative framework is adopted to motivate new rewrites, leading to dynamic update of rewrite vocabulary.
    \item The traditional rewrite task is formulated as comprehensive query process including successive tasks of query understanding, correction, clarification and rewrite in generation, and multi-tasks post-training.
\end{itemize}

\section{Related Works}
Query rewrite, or query expansion has long been a crucial technique to enhance the efficiency and accuracy of information retrieval and database querying. It serves as an important component for search in both general Web search (Google, Baidu) and e-Commercial search (Taobao, Meituan Delivery). 
The methods for query rewrite can be broadly categorized into two phases: pre-LLM methods and LLM based methods.

Before the advent of LLMs, query rewrite techniques primarily relied on rule-based systems and traditional machine learning approaches. These methods often involved manually crafted rules or heuristics to transform user queries into more effective or efficient forms. 
One of the early approaches to query rewrite involved using thesauri and ontologies to expand queries with synonyms and related terms. \cite{voorhees1994query} explored the use of WordNet, a lexical database, to automatically expand user queries with semantically related terms.
Another significant pre-LLM approach was the application of statistical machine translation (SMT) models to query rewriting. \cite{berger2017information} proposed using SMT techniques to translate user queries into a language model that better represents the document space. This method treated query rewriting as a translation problem, leveraging co-occurrence statistics from large corpora to infer possible rewrites that could enhance retrieval accuracy.
\cite{jones2006generating} investigated the use of query logs to identify common reformulations that led to successful search outcomes.
\cite{cui2002probabilistic} proposed a method that utilized query expansion through the use of frequently co-occurring terms in large text corpora to improve search engine performance. Similarly, \cite{baeza2004query} explored query transformations using statistical translation models to enhance information retrieval systems. 
These approaches, while effective to some extent, often required significant manual effort to maintain and update the rule sets and were limited by their inability to generalize well across diverse query contexts.

With the introduction of LLMs, such as BERT\cite{devlin2019bert} and GPT\cite{radford2018gpt1}, query rewrite methods have seen significant advancements. These models leverage deep learning and vast amounts of training data to automatically learn complex patterns and relationships in language, enabling more sophisticated and context-aware query transformations. 
For example, \cite{nogueira2019passage} demonstrated the use of BERT for query reformulation in search engines, showing substantial improvements in retrieval performance. 
\cite{peng2024large} pioneer to deploy LLM-based query rewrite in e-Commercial search system by training a LLM for query rewrite with multi-instruction supervised fine tuning, offline feedback, and objective alignment.
\cite{wang2024one} models the query rewrite as an end-to-end keyword generation task based on LLM by multi-match prompt tuning and prefix tree-based constrained beam search in generation. Then it employs feedback tuning for LLM post training.
These LLM-based methods heavily reply on prompt tuning and post-training of LLMs in a one-time generation, without consideration of rewrites update.

\section{Method}
\method{} is organized as an iteration. As shown in Fig.\ref{fig:overall}, we divide the whole procedure into following three stages: 1) Initialization \& Generation 2) Online Signal Collection 3) LLM Post Training for query rewrite.

In each iteration: rewrites are firstly generated based on open-source / commercial LLMs or post-trained LLMs in previous iteration. The generated rewrites are then de-duplicated with online rewrites and deployed online, which served to retrieve items for the corresponding query. The clicked / purchased items indicate positive feedback from user, leading to labeling positive for the rewrites. These positive rewrites are collected, together with auxiliary tasks such as Quality, Relevance for post-training.

\subsection{Initialization \& Generation}\label{sec:rewrite-generation}
This stage generates rewrites given queries by LLMs, including open-source / commercial LLMs in the very beginning iteration and post-trained LLMs in subsequent iterations.

Prior to generation, we separate queries into three categories based on their frequency proportion: high frequency, mid frequency and tail. 
High frequency queries includes most frequent search queries in our system, such as `beef noodles', `snack'. These queries represent a majority of user preferences and contribute to the largest Gross Merchandise Volume (GMV). Besides, the number of high frequency queries is limited.
Mid frequency queries normally comprise specific search intention, such as ``Luosifen'', ``Xijiade''\footnote{Xijiade Dumplings is a well-known Chinese restaurant chain specializing in dumplings and traditional northeastern Chinese cuisine}. Its volume is lower while the number increases compared to the high frequency counterpart.
Tail queries are paid less attention before. Basically it includes typo of user, synonyms, particular local food / restaurant, equivocal search intention, natural language and etc., such as ``KFC'', ``Fat Uncle'', ``Extra Spicy'', ``What is suitable to eat when having a cold?''. 
It is difficult to infer user's intention based on tail queries. 

We further set up 5 rewrite directions for comprehensive rewrites: 1) Key Word Extraction 2) Correction 3) Alias \& Synonyms 4) Main Dish 5) Low Relevance. The exact definition, positive/negative examples are listed in App.\ref{app:rewrite-direction}.

During rewrite generation by LLMs, prompt is designed as followed: 
We set different prompts for queries in different categories, with explanation of the category and rewrite emphasis. CoT and RAG are also employed in this process.

\subsubsection{Chain-of-Thoughs (CoT)}
CoT is utilized to formulate task of query rewrite to comprehensive query understanding and process.
We instruct LLM to first determine the meaning of the search query: Whether it is unambiguous. If not, the query should be corrected or rephrased.
Then, LLM is required to determine the search intent: whether user tend to match cuisine names or restaurant names under the query.
Finally, LLM generates rewrites under the information inferred by itself.

\subsubsection{Retrieval-Augmented Generation (RAG)}
Generation process is further enhanced by incorporating relevant information. Specially, for each query, the associate restaurants and cuisine with interaction are added to the prompt.

Finally we let LLMs to choose suitable rewrite directions and conduct final rewrite. 
Prompt for query generation and examples is listed in Appendix \ref{app:rewrite-prompts}. 

\subsection{Online Signal Collection}\label{sec:signal-collection}
Most previous rewrite methods find it difficult to determine whether a rewrite is ``good'' or not automatically. Inspection method includes manual / GPT labeling for selected rewrites, bringing high-cost manpower efforts and indirect connection to final results.
In \method{}, we proposed to use real-world serving system to determine the quality of rewrites: the rewrites contributes to click / purchase action from users are collected and regarded as ``good'' rewrites. 

Specially, during serving in our online system, input query is rewritten to multiple rewrites. Each rewrite is used to retrieve candidate items for ranking and expose. 
Besides rewrite retrieval, the system comprises other retrieval methods, including origin query retrieval, embedding retrieval, user-to-item (U2I) retrieval and etc., whose retrieved items are mixed and de-duplicated. 
% We specially mark the item: 1) Retrieved by rewrites (retrieval by other methods as well) 2) Only retrieved by rewrites. For the marked 
We specially mark the items that are retrieval (only) by rewrite retrieval. 
During expose, the clicked item that are only retrieved by rewrites is regarded as ``Level-1'' positive signal to the rewrite, meaning that the rewrite provides unique contribution to the click action. For the clicked item retrieved by multiple method include the rewrite, we regard it as ``Level-2'' positive signal, where the rewrite satisfies the search intention, however without increment to the search results.
``Level-1/2'' rewrites are collected and utilized as positive label for latter process. 

\subsection{Multi-Task Post Training for Query Rewrite}\label{sec:llm-post-training}
Public LLM are trained for general usage including reasoning, conversion and etc. \method{} aims at training a specific LLM for query rewrite, based on the understanding of search query and matching of candidate items in our online system. 
Specially, the rewrite is supposed to 1) precisely represent user's intention in Meituan, 2) reason to specific item for query in natural language format, 3) relevance between query and 4) capture inherent user's interest.

To achieve these goals, three objectives and corresponding loss are established during stage of post training: 1) Rewrite Generation 2) Rewrite Quality 3) Relevance. 
% Specially, given query and corresponding positive rewrites from Sec.\ref{sec:signal-collection}, we firstly inference query's intention using our query understanding module, relevance between query and rewrites using relevance module \footnote{GPT-4 is actually used in experiments}. 
% Then we construct a prompt based on Sec.\ref{sec:rewite-generation} with similar CoT, leaving positive rewrites and their corresponding intention, relevance score masked for learning.
\subsubsection{Rewrite Generation}\label{sec:llm-post-training:rewrite-generation}
Rewrite generation serves as the main task in post training. LLM is trained to produce rewrites to match user's intention and interest.
With the collected rewrites with positive signal, instead of training LLM simply to generate them, we further rephrase the generation task using CoT by dividing it into several subtasks: 
Given the query in training samples, We first instruct LLM to tell whether the query is typo, i.e. wrongly written words. 
Then the LLM distinguishes the intention of query from choices of cuisine, restaurant and neither.
Finally, LLM is required to generate rewrites.
The prompt is arranged by concatenation of the three above mentioned subtasks. 
User input is consist of query and its corresponding names of most frequently clicked cuisine and restaurant.
Assistant output comprises the label of the subtasks: 
We use other LLMs (such as GPT-4o) to tell whether the query is typo or not in advance. 
Intention of query is inferred with our query understanding module. 
The rewrites collected in Sec.\ref{sec:signal-collection} are utilized. 
The training instruction shares the same with prompt in Sec.\ref{sec:rewrite-generation}. Besides, the content of assistant output is replaced by the positive rewrites.

\subsubsection{Rewrite Quality}\label{sec:rewrite-quality}
With the generated rewrites by LLM, we need to distinguish whether the rewrite are suitable for the system or not. Though rewrites with positive signals are trained in Sec.\ref{sec:llm-post-training:rewrite-generation}, rewrites without signals are unseen for LLM, leading to repeated occurrence in LLM rewrite generation. 
To endow LLM with the ability to generate ``good'' rewrites instead of ``bad'', we establish a classification task for LLM: During post-training, LLM is instructed to produce ``good'' or ``bad'' labels for a input pair of query and corresponding multiple rewrites. 

Similar with prompt in Sec.\ref{sec:llm-post-training:rewrite-generation}, we instruct LLM of the criterion of quality in instruction. 
Then user input is consist of query, cuisine and restaurant that are most interacted and rewrites generated from previous iteration. 
Assistant output comprises the labels, which are generated based on the online signal in Sec.\ref{sec:signal-collection} and auxiliary labeling using other LLMs (such as GPT-4o): the rewrites with positive signal is regarded automatically as ``good''. The label of remaining rewrites are inferred with other LLM. 
Training sample is listed in Appendix \ref{app:post-training:rewrite-quality}.

\subsubsection{Relevance}
Relevance between query and search results is a main constraint in search system. 
To fulfill the restriction, the generated rewrite is supposed to match high relevance with query. 

We instruct LLM to be aware of the relevance between query and rewrite. 
Specially, the instruction is a detailed description of relevance inference procedure. 
User input is composed of query, cuisine and restaurant under the query. 
Assistant output is the label of relevance including high/low/non relevance.
Similar with Sec.\ref{sec:rewrite-quality}, we use our relevance measuring module for labeling query and rewrites into the three levels. 
Besides, a auxiliary LLM is utilized to perform labeling. We choose the training samples with agreement.
Training sample is listed in Appendix \ref{app:post-training:relevance}.

As shown in Alg.\ref{alg:overall}: After post training of LLM, we use the trained LLM for generation using prompt in Sec.\ref{sec:rewrite-generation}, whose output rewrites are fed online for signal collection.
In each iteration, we add more queries and modify prompts based on online performance.
After iterations of collection and post-training, the generated rewrites tend to fulfill the objectives and lead to better online performance. 
\begin{algorithm}
\caption{\method{}: An Iterative Procedure for Rewrite Generation}\label{alg:overall}
\begin{algorithmic}[1]
% \Require $n \geq 0$
% \Ensure $y = x^n$
% \State $y \gets 1$
% \State $X \gets x$
% \State $N \gets n$
\State $\text{ite}=0$
\While{True}
% \If{$N$ is even}
%     \State $X \gets X \times X$
%     \State $N \gets \frac{N}{2}$  \Comment{This is a comment}
% \ElsIf{$N$ is odd}
%     \State $y \gets y \times X$
%     \State $N \gets N - 1$
% \EndIf
\State Generate query rewrite using public LLMs using prompt in Sec.\ref{sec:rewrite-generation}.
\If{$\text{ite}>0$}
    Generate query rewrite using post-trained LLM using prompt in Sec.\ref{sec:rewrite-generation}.
\EndIf
\State Rewrites de-duplicated and deployed.
\State Collect positive rewrites.
\State LLM post-training with multi-tasks in Sec.\ref{sec:llm-post-training}
\State $\text{ite}+=1$
\EndWhile
\State Collected positive rewrites are deployed.
\end{algorithmic}
% \caption{Algorithm description of \method{}.}
\end{algorithm}

\section{Experiments}
Experiments of \method{} mainly answer the following questions:
\begin{itemize}
    \item Whether rewrites generated by \method{} contributes to online improvement in our real-world service, i.e. search in Meituan delivery.
    \item Whether \method{} generates better rewrites than public LLMs in the domain of Meituan delivery, especially in searching of cuisine and restaurant.  
    \item Whether the iterative framework produces better rewrites automatically. 
\end{itemize}
To address first concerns, we first conduct online A/B test to verify the effectiveness of \method{} in real-world scenario. 
Specially, in retrieval phrase of Meituan Delivery's search, \method{} is added as another channel for retrieval. The retrieved items of \method{}, together with items from other channels are truncated, de-duplicated and merged for phrase of rank.
Finally, the ranked items are exposed to user. 
Online performance for search system with and without \method{} are recorded and compared as A/B test.

% Besides, for further verifying the performance of rewrites before deploy and providing guidance of the rewrite refinement, we design and establish offline evaluation, including 3 sets of experiments.

The second question serves as an pre-deployment validation for the effectiveness of \method{}, providing guidance for the development of rewrite generation.
It is difficult to directly measure the online metric before deployment, offline but related metrics are established.
It is subjective to judge the quality of generated content \cite{guo2023close}.
However, rewrites contain meaningful representation and serves in a specific task: reformulate user input to similar query for available and satisfied candidates matching.
Given an input, it is possible to enumerate suitable rewrites as ground-truth, measuring the relevance between generated rewrites and input to ensure consistent of search intention and estimate the efficiency of rewrites by simulation of online serving process.
Based on these assumption, 3 sets of offline experiments are set to evaluate the performance of rewrites.

Specially, a benchmark containing queries and multiple corresponding rewrites as ground-truth is prepared. 
During offline experiments, after rewrites are generated, the coverage rate of generated rewrites by different methods w.r.t ground-truth is calculated as the ``precision'' of methods.
This metric measures whether the method is able to produce ``good'' rewrites that are labeled manually by common sense.
Furthermore, A relevance score between generated rewrites and queries is calculated by our relevance module, which serves as the metric for measuring the consistence of search intention during rewrite process.
This metric is named as ``Relevance''.

Another benchmark is established to simulate the online procedure.
Specially, given a user query, corresponding restaurants and cuisines (with interaction labels) serve as candidates for retrieval. Embedding-based retrieval is used: query and title of candidates are embedded, the inner product between query and candidates is calculated as score for retrieval. A higher score represents higher probability for exposure. 
A good query is supposed to generate higher score for restaurants and cuisines with interaction labels.
During rewrite experiments, we replace origin query with rewrites generated from different methods for embedding and retrieval.
Though real-world search system comprises sophisticated modules, the offline simulation is able to roughly measure the efficiency performance of rewrites: whether the generated rewrite retrieve more interacted items than original query. 
This experiment measure the ``Efficiency'' of rewrites in an offline environment by simulating the real-world procedure of rewrite retrieval.

The third concern can be partial answered by the experiments above mentioned. 
Besides, we focus more on the rewrites generated by post trained LLM during each iteration: whether \method{} continuously motivates new rewrites.
Compared with the LLM based query rewrite methods \cite{peng2024large}, \cite{wang2024one} where one-step rewrite generation is conducted, \method{} employs iterative generation of rewrites. 
The proportion and performance newly generated is measured.

\subsection{Online Experiments}
Specially, a new recall channel is set up for the rewrites generated after iterations of rewrite generation.
For a user query, 10 arbitrary corresponding rewrites are sampled to retrieve items from candidate sets\footnote{Normally, the candidate sets for recall are the available restaurants and cuisine according to user's current location.}. 
These retrieved items, together with the counterparts in query text/embedding recall,  user historical behavior recall and etc, are de-duplicated for ranking and exposure.
Finally, user's purchase information is reported for comparison. Specially,  
we select order volume, conversion rate from exposure to order per expose (PV\_CXR) / query (QV\_CXR) / user (UV\_CXR). 
% (UV\_CXR: Conversion rate from exposure to purchase per user; UV\_RPM: Revenue Per Mille for every exposed user) are utilized to measure the performance of the overall system. 
We set a search system without \method{} as baseline and system with \method{} as target. Performance metrics 
(Order Volume, PV\_CXR, QV\_CXR, UV\_CXR) 
of target above baseline is reported as the improvement brought by \method{}.
Besides, we set multiple baseline systems, whose differences are recorded as fluctuation.

\method{} is deployed online for 7 days. As shown in Table \ref{tab:abtest}: during experiment, consistent improvement is observed for \method{}.   

\begin{table}[!htbp]
    \centering
    \resizebox{\linewidth}{!}{
    \begin{tabular}{c|c|c|c|c}
        \hline \hline
        Experiments & Order Volume(\%) & PV\_CXR(\%) & QV\_CXR(\%) & UV\_CXR(\%) \\ \hline
        AA & 0.04 & 0.02 & 0.09 & 0.03 \\  \hline
        AB(\method{}) & 0.27 & 0.34 & 0.27 & 0.21 \\  \hline  \hline
    \end{tabular}
    }
    \caption{Online experiment of \method{}.}
    \label{tab:abtest}
\end{table}

\subsection{Offline Experiments}\label{sec:offline-experiments}
Online experiment evaluates the overall performance of \method{}. 
% We dissect the working procedure of rewrites
To examine the performance of rewrites in detail and foresee online performance before deployment, % 3 offline experiments are established.
we establish $3$ offline experiments and $2$ corresponding benchmarks.

Specially, benchmark I contains $3597$ queries and average $5.19$ positive rewrites for each query. 
The positive rewrites serve as ground truth rewrites that are supposed to be generated, which is generated by GPT-4o and examined manually for correctness.
These rewrites never appear in generation or post-training phase to avoid knowledge leakage.

Benchmark II contains $10000$ queries and corresponding recall candidate set with average volume of $500$, the candidate set comprises the name of restaurants within the location constraint for the request. Statistics of the benchmarks is listed in Table \ref{tab:benchmark}.

\begin{table}[!htbp]
    \centering
    \resizebox{\linewidth}{!}{
    \begin{tabular}{c|c|c|c|c}
    \hline \hline
        Benchmark ID & $\#$Query & Label Content & $\#$Candidates & $\#$Label \\ \hline
        I & $3597$ & Positive rewrites & - & $5.19$ \\ \hline
        II & $10000$ & Clicked items & $500$ & 2.3 \\ \hline \hline
    \end{tabular}}
    \caption{Benchmark specification and statistics. $\#$ represents the number of selected content.}
    \label{tab:benchmark}
\end{table}
For evaluation methods, besides \method{}, we use GPT-4o, DeepSeek-v2 as close source baselines; Qwen2.5 as open source baseline.
\method{} is built on Qwen2.5, with CoT \& RAG in generation and multi-task post-training in an iterative training framework.
Ablation study is also conducted on these components.

\begin{table}[!htbp]
\centering
\resizebox{\linewidth}{!}{
    \begin{tabular}{r|c|c|c|c|c}
    \hline \hline
    \multirow{2}{*}{Method} & \multirow{2}{*}{Precision} & \multirow{2}{*}{Relevance} & \multicolumn{3}{c}{Efficiency} \\ \cline{4-6}
        &   &    & Top1 & Top5 & Top10 \\ \hline
    GPT-4o & 0.2131 & 0.1901 & 0.0033 & 0.0231 & 0.0534 \\ \hline
    DeepSeek-v2 & 0.2245 & 0.2173 & 0.0043 & 0.0247 & 0.0567 \\ \hline
    Qwen2.5 & 0.1589 & 0.1280 & 0.0034 & 0.0219 & 0.0522 \\ \hline
    Qwen2.5 + PT(RG) & 0.3649 & 0.4503 & 0.0067 & 0.0458 & 0.0971 \\ \hline
    Qwen2.5 + PT(RG) + RAG & 0.4752 & 0.4149 & 0.0057 & 0.0413 & 0.0943 \\ \hline
    % Qwen2.5 + PT + RAG + CoT & & & & \\ \hline
    % Qwen2.5 + PT + RAG + CoT + Iter (\method{}) & & & & \\ \hline
    % \method{} w.o. RAG & 0.4752 & & & \\ \hline
    \method{} w.o. Iter\&PT(Rel) & 0.4807 & 0.4110 & 0.0063 & 0.0442 & 0.0952\\ \hline
    \method{} w.o. Iter & 0.4833 & \textbf{0.4149} & \textbf{0.0086} & \textbf{0.0465} & 0.0955\\ \hline
    \method{} & \textbf{0.5040} & 0.3362 & 0.0058 & 0.0364 & \textbf{0.1017} \\ \hline \hline
    \end{tabular}
}
\caption{Overall offline experiment results of \method{} compared with baselines. PT stands for Multi-task post training; PT(RG) means post training with only Rewrite Generation. PT(Rel) represents post training with only Relevance.}
\label{tab:offline}
\end{table}

\subsubsection{Precision Generation of Rewrites}
For each query, we evaluate the precision of generated rewrites: for each query in the benchmark I, we generate rewrites using different methods. 
% During generation, we use the same prompt for all methods for fairness.
% Then the coverage of generated rewrite with respect to the positive rewrites are calculated:
Then we calculate the coverage portion of positive rewrites:
A good rewrite LLM is able to generate all positive rewrites, leading to higher coverage. 

As shown in Table \ref{tab:offline}, open/close source LLMs with general usage demonstrate apparent degeneration than LLMs post trained with domain data. 
A single rewrite generation training task on Qwen2.5 improves by large margin (0.3649 v.s. 0.1589), showing that domain knowledge contributes significantly. 
Besides, RAG  assistants a post-trained LLMs to better performance. 
By adding multi-task post training and CoT/RAG in generation, \method{} without iteration leads to higher precision.
Finally, iterative framework further boost performance to 0.5040, meaning that half of the reserved rewrites can be generated.

\subsubsection{Relevance of Rewrites}\label{sec:exp:relevance}
In this experiment, relevance of rewrites with respect to query is evaluated. 
%  all generated rewrites' relevance with respect to query are inferenced using our own relevance module. 
We use our own relevance module to inference the relevance scores.
The module outputs prediction for a pair of query and rewrite: high-relevance, low-relevance and non-relevance. The portion high is reported as the capacity of LLM to generate relevant rewrites.

As shown in Table \ref{tab:offline}, general LLMs demonstrate poor performance before post training, showing that though public LLMs perform well in conversation, reasoning, relevance knowledge in e-commercial domain is lacking.
Besides, post-training in rewrite generation leads to significant improvement in relevance.
It is found that an ablation experiment on \method{} without Relevance demonstrates degradation in both precision and relevance metric.
However, iterative training on Relevance does not leads to improvement in relevance metric. We conjecture that Relevance is over fitted during multiple training.

\subsubsection{Efficiency of Rewrites}\label{sec:exp:efficiency}
Finally, we use the benchmark II to simulate the online procedure and estimate the efficiency of rewrites.
Similarly, we first generate rewrites for all methods. 
Then the rewrites, together with the candidate items are embedded to numerical representation. 
We use inner product between rewrites and items for scoring to simulate the online recall and ranking process.
The items with user's click are regarded as positive samples, whose scores are supposed to be higher than other samples. 
We use recall@$K$ to measure the performance of the offline simulation: for the $K$ items with highest score, the number of positive samples within is recorded as numerator while the number of positive samples is recorded as denominator.

As shown in Table \ref{tab:offline}, general LLMs show inferior performance in retrieval efficiency, compared with post trained LLMs.
Post training with Rewrite Generation (Qwen2.5+PT(RG)) contributes significantly to improvement. Some auxiliary tasks such as Relevance and RAG may lead to degradation.
Finally, all multi tasks and prompt techniques integrated in \method{} reaches to best performance in all metrics of Efficiency.

\subsection{New Rewrites in Iterations}
During the procedure of \method{}, the initial set of rewrites is generated by public LLMs (GPT-4o, DeepSeek-v2).
Starting from the first iteration, the post trained LLM in \method{} generates rewrites.
Besides, the previous rewrites is filtered by Stage 2 (Online Signal Collection) in last iteration, leaving only the positive rewrite in the current iteration.
These rewrites are de-duplicated for Stage 2. 
To examine the performance of newly generated rewrites in post trained LLM, we only use the unique rewrites for experiments:
The portion of these rewrites is recorded, demonstrating the contribution of post trained LLM in motivating rewrites.
Furthermore, similar with the experiments in Sec.\ref{sec:offline-experiments}, we compare the offline performance of the unique rewrite.
Considering only the unique rewrites is compared, most rewrites are covered by the common parts, experiment of Precision is omitted.
During the experiment, we use three times of iterations in \method{}.

\begin{table}[!htbp]
    \centering
    \resizebox{\linewidth}{!}{
    \begin{tabular}{c|c|c|c|c}
     \hline \hline
    \multirow{2}{*}{Iteration Index} & \multirow{2}{*}{Relevance} & \multicolumn{3}{c}{Efficiency} \\ \cline{3-5}
        &        & Top1  & Top5    & Top10 \\ \hline
    1   & 0.3466 & 0.0034 & 0.0412 & 0.0767 \\ \hline
    2   & 0.4021 & 0.0045 & 0.0308 & 0.0923 \\ \hline
    3   & 0.3677 & 0.0054 & 0.0428 & 0.0917 \\ 
    \hline \hline
    \end{tabular}}
    \caption{Offline experiments (Relevance \& Performance) of unique rewrites in three iterations.}
    \label{tab:exp:unique-new-rewrite}
\end{table}

\begin{figure}[!htbp]
    \centering
    \includegraphics[width=\linewidth]{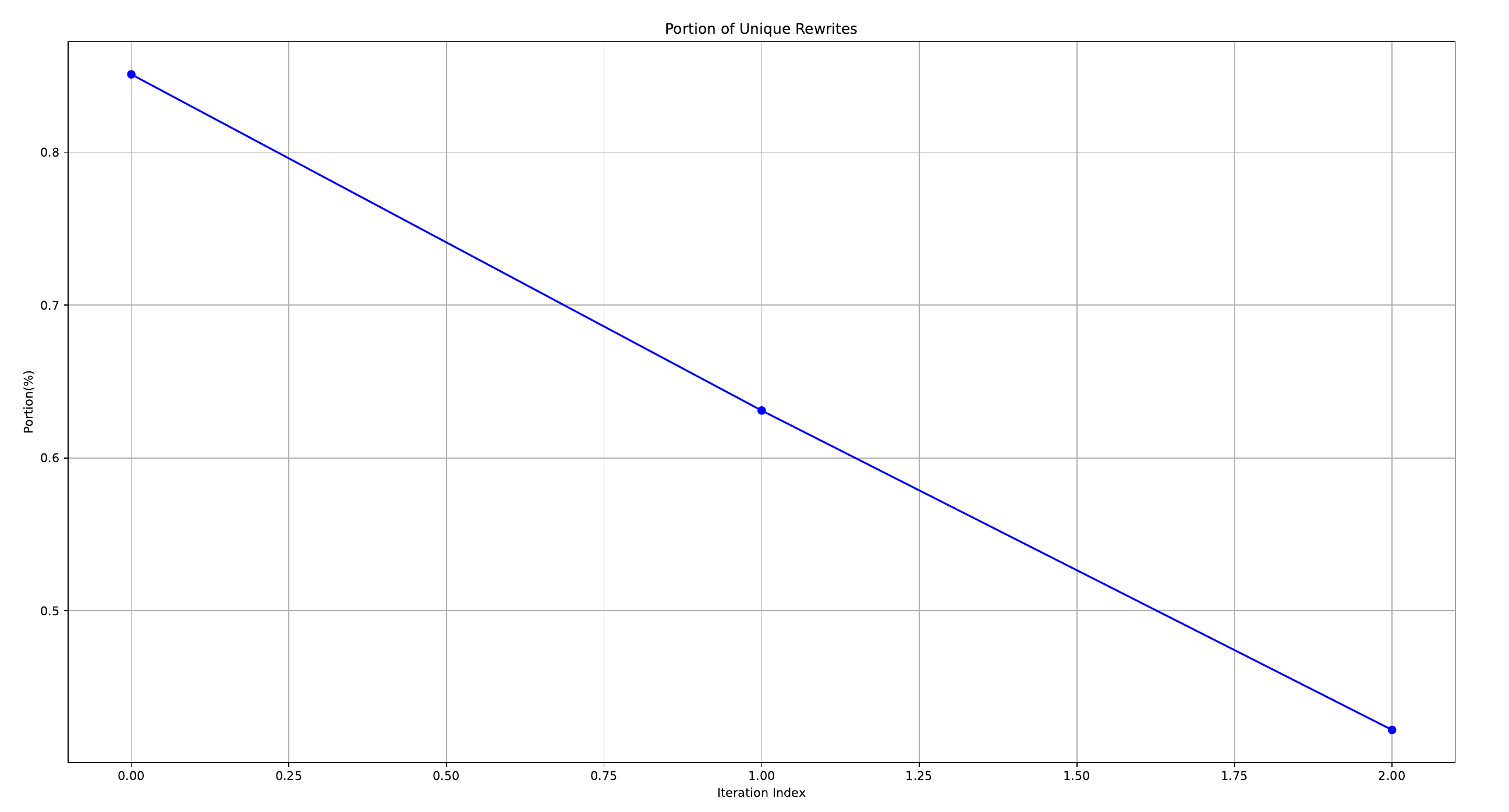}
    \caption{Portion of unique rewrites generated by \method{} during iterations}
    \label{fig:unique-rewrite-portion}
\end{figure}

\subsubsection{Portion of New Rewrites}
During iteration, the rewrites generated by post trained LLM in Stage 1, together with the filtered rewrites in Stage 2 (previous iteration) are de-duplicated and fed for online signal collection.
The unique rewrites provided by post trained LLM represents the contributed rewrites of \method{} for the iteration. 

The portion of unique rewrites during the three iterations is demonstrates in Fig.\ref{fig:unique-rewrite-portion}. 
In iteration 0, rewrites are generated by public LLMs. New rewrites motivate in iteration 1, where new rewrites occupy 0.852\% over the all rewrites used in this iteration.
Though it is of minor occupation, it actually contributes new information for the search system. 

As iteration progresses, portion of new rewrites keep decreasing, showing that the overall available rewrites for the search system achieve stability in iterations.

\subsubsection{Performance of New Rewrites}
We further evaluate the offline performance for the new rewrites during iterations.
Specially, after rewrite generations of post trained LLM, the rewrites are de-duplicated with the positive rewrites. 
The remaining rewrites' Relevance and Efficiency are evaluated similarly in Sec.\ref{sec:exp:relevance} and Sec.\ref{sec:exp:efficiency}.

As shown in Table \ref{tab:exp:unique-new-rewrite}, during iteration progresses, metrics of relevance and efficiency maintain similar results. 
It demonstrates that the performance of new rewrites does not change dramatically. The performance of post trained LLM is stable during training in multiple iterations.
The data also resemble corresponding results in Table \ref{tab:offline}. It shows that the new rewrites' performance is similar with the positive rewrites.

\section{Conclusion}
In this paper, we proposed \method{}, an iterative framework to generate rewrites by meticulous prompt tuning and post training of LLMs.
Specially, in each iteration, the prompt formulates query rewrite into a query understanding and process task by CoT and incorporate associated interacted restaurant and cuisine of the query as RAG. 
The generated rewrites are deployed to collect online feedback. The positive rewrites is reserved for their leading restaurant and cuisine is clicked by users.
The collected rewrites, together with the relevance labels are utilized in a multi-task post training of LLM, which motivates new rewrites in next iteration to supplement deployment for feedback collection.

\clearpage
\section{Reference}
\bibliographystyle{plain}
\bibliography{references}

\clearpage
\section{Appendix}
\subsection{Rewrite Direction}\label{app:rewrite-direction}
\begin{table}[!htbp]
\centering
\resizebox{\linewidth}{!}{
\begin{tabular}{p{0.25\linewidth}|p{0.25\linewidth}|p{0.25\linewidth}|p{0.25\linewidth}} \hline \hline
    Direction & Definition & Positive Cases & Negative Cases \\ \hline
    Key Word Extraction & Extract the core content from the query, which must be entirely contained within the query & Animal cream birthday cake $\to$ cake,  birthday cake & Linwei's Chuan $\to$ barbecue \\ \hline
    Correction & Rewrite the query to use correct wording. & Wontom $\to$ Wonton & HK $\to$ HK style Cafe \\ \hline
    Alias \& Synonyms & Rewrite the query to commonly used wording. & kfc $\to$ Kentucky Fried Chicken & kfc $\to$ McDonald
    \\ \hline
    Main Dish & Rewrite the query to cuisine that are short, common, and highly related to the original query; cannot directly use the provided query. & Daifuku Spicy Kimchi $\to$ Kimchi Soup, Spicy Pickled Cabbage & Daifuku Spicy Kimchi $\to$ Bibimbap \\ \hline
    Low Relevance & Rewrite the query to similar cuisine. & Big Pork Bone $\to$ Steak & Big Pork Bone $\to$ Northeast Braised Pork Bones  \\ \hline \hline
\end{tabular}}
\caption{Rewrite direction}
\label{tab:rewrite-direction}
\end{table}

\subsection{Prompts for Rewrite Generation}\label{app:rewrite-prompts}
\begin{tcolorbox}[title = {Prompt for Rewrite Generation for Tail Query}]
\texttt{Instruction}: 

You are a query analysis expert for a food delivery platform.
You are provided with user search queries, along with the standard names of the restaurant and cuisine that have historically received the most clicks and purchases for those queries. Please:

1) First, analyze the meaning of the query. If there is an input error, typo in the query, please correct it. If no correction is needed, output ``None'';

2) Then, determine the search intent. Do user tend to match cuisine names or restaurant names under the query? Choose one from ``Cuisine'', ``Restaurant'', or ``Neither'';

3) After that, provide N query rewrites for the original query, and output them in order of rewrite efficiency from high to low. The query rewrite can include the following types:

- Key word extraction: Extract the core content from the query, which must be fully contained within the query;

- Alias \& Synonyms: To search for more similar dishes, the terms should be short and common, and should not directly use the provided dish names;

- Main Dish: For example, ``burger'', ``cake'', ``noodles''. Note that dish category terms should be specific and not too general or vague;

- Related cuisine (Low Relevance): The rewrite should be short, common, and highly related to the original query, and should not directly use the provided restaurant and cuisine names;

Finally, output only in the following format: {``Query meaning'': ``Summarize in less than 30 words'', ``Correction'': ``Correction result'', ``Search intent'': ``Cuisine/Restaurant/Neither'', ``Rewrite'': ``rewrite1, rewrite2, ...''}.

\texttt{User}:

Query:\{\} 

Associated restaurant/Cuisine:\{\}

Query Explanation: It is highly possible that the query contain type, synonyms,  particular local food \/ restaurant, equivocal search intention, natural language and etc, leading to a low appearance frequency. Try to infer the actual meaning of the query.

\texttt{Assistant}: 

Output: \{\}
\end{tcolorbox}

Different categories of query applies to different query explanation:

\begin{tcolorbox}[title = {Query explanation for Query with High Frequency}]
This query is commonly used, it may represent a board category or common cuisine. Try to extract the key word and find cuisine that are with low relevance for interest exploration.
\end{tcolorbox}

\begin{tcolorbox}[title = {Query explanation for Query with Mid Frequency}]
This query may be a local food or specific brand name. Try to find out the main dish the query contain.
\end{tcolorbox}

\subsection{Post-Training}

\subsubsection{Rewrite Quality}\label{app:post-training:rewrite-quality}
Training Sample for Rewrite Quality:
\begin{tcolorbox}[title = {Training Sample for Rewrite Quality}]
\texttt{Instruction}:

You are a query rewrite evaluation expert for a food delivery platform. Your task is to determine whether a given query rewrite is a good rewrite based on the user's input query, names of the restaurant and cuisine with the most historical clicks and purchases for that query.

A good query rewrite should ensure strong relevance to the original query term while also enabling the incremental recall of more dishes that users might click and order.

You will be provided with a group of two queries and corresponding information each time. Please evaluate them in sequence and respond with ``Yes'' or ``No''.

\texttt{User}:

Query1: \{Query\}; Associated restaurant/Cuisine1:\{\}

Query2: \{Query\}; Associated restaurant/Cuisine2:\{\}

\texttt{Assistant}:

Output: \{1.Yes. 2.No\}
\end{tcolorbox}

\subsubsection{Relevance}\label{app:post-training:relevance}
Training Sample for Relevance:
\begin{tcolorbox}[title = {Training Sample for Relevance}]
\texttt{Instruction}:

You are an expert in a food delivery platform. Your task is to determine the relevance level of user search query to the products based on the following scoring criteria, categorized into three levels: High, Low, and None. You will be given user search query, names of restaurant and cuisine.

First, you need to analyze the user's search intent based on the user query. Extract the type of restaurant and cuisine the user wants to buy and any specific attributes they require (such as ingredients, taste, preparation method, size).

High Relevance: If the restaurant and cuisine is of the type the user wants to buy, and the user does not have specific attribute requirements, or the restaurant and cuisine meets all the user's attribute requirements, then the restaurant and cuisine fully satisfies the user's needs.

Low Relevance: If the restaurant and cuisine is of the type of cuisine the user wants to buy but does not fully meet the user's attribute requirements; or if the user does not have specific attribute requirements, and although the restaurant and cuisine is not of the type of dish the user wants to buy, it can still meet the user's needs based on the intended use.

No Relevance: If the restaurant and cuisine is not of the type of dish the user wants to buy and cannot meet the user's needs based on the intended use, then the restaurant and cuisine does not satisfy the user's requirements.

You will be provided with three groups each time. Please evaluate them in sequence and respond with the relevance level: ``High'', ``Low'', or ``None''.

\texttt{User}:

Query1: \{Query\}; Restaurant1: \{\}; Cuisine1: \{\}

Query2: \{Query\}; Restaurant1: \{\}; Cuisine2: \{\}

Query3: \{Query\}; Restaurant1: \{\}; Cuisine3: \{\}

\texttt{Assistant}:

Output: \{1.Low. 2.High 3. None\}
\end{tcolorbox}

\end{document}